\documentclass{ws-ijmpd}
\usepackage[super,compress]{cite}
\begin{document}

\markboth{Romero, G.E., Sotomayor, P.}
{Population III Microquasars}

%
\catchline{}{}{}{}{}
%

\title{POPULATION III MICROQUASARS}

\author{ROMERO, GUSTAVO E.}

\address{Instituto Argentino de Radioastronom\'ia (IAR, CCT La Plata, CONICET)\\
C.C.5, (1984) Villa Elisa, Buenos Aires, Argentina\\
        Facultad de Ciencias Astron\'omicas y Geof\'isicas, Universidad Nacional de La Plata\\
Paseo del Bosque s/n, 1900, La Plata, Argentina\\
gustavo.esteban.romero@gmail.com}

\author{SOTOMAYOR CHECA, P.\footnote{pablosotomayor.fcag@gmail.com}}

\address{Facultad de Ciencias Astron\'omicas y Geof\'isicas, Universidad Nacional de La Plata\\
Paseo del Bosque s/n, 1900, La Plata, Argentina\\
pablosotomayor.fcag@gmail.com}

\maketitle

\begin{history}
\received{Day Month Year}
\revised{Day Month Year}
\end{history}

\begin{abstract}
We present the first results obtained in the elaboration of a complete model of a microquasar where the donor star is from Population III. These stars do not produce stellar winds so we consider that the mass loss is due exclusively to matter overflowing the Roche lobe towards the compact object, a maximally rotating black hole. The rate of accretion is extremely super-Eddington, with an intense mass loss from the system in the form of winds and jets. We calculate the relativistic particle content of the jet and the corresponding spectral energy distribution (SED) considering a lepto-hadronic model. Prospects for the cosmological implications of these objects are briefly discussed.
\end{abstract}


\ccode{PACS numbers:52.27.Ny; 97.10.Gz; 97.80.Jp; 98.80.−k}


\section{Introduction}	

The first stars are currently considered to be the main sources of reionization of the Universe. Because of to their temperature, these stars emit most of their radiation in the ultraviolet region of the electromagnetic spectrum. However, the fraction of UV photons that escapes from the first galaxies to produce and maintain ionization in regions far from the intergalactic medium is uncertain. Recent observations from the Hubble Space Telescope suggest that ultraviolet radiation from the most distant galaxies detected so far is not sufficient to ionize the IGM on large scales. To solve this problem it is necessary to explore other possible sources of reionization.\par
The first accreting binary systems would have played an important role in the reionization of the Universe\cite{Mirabel_BH_2011}, ionizing in largea volumes of space in regions of low density IGM. Cosmic rays produced in Population III microquasars jets are another possible source of reionization\cite{Tueros_CR_2014}. These suggestions, although attractive and energetically consistent, are not supported by a specific model of Population III microquasar model. In this work our primary goal is to start the elaboration of a complete model of these objects, in order to make quantitative predictions about their effects. As a secondary objective we intend to make realistic predictions of the production of radiation and cosmic rays that will be injected into the intergalactic medium, which are important for their ionizing power for later studies on the early Universe.

\section{Model}

\subsection{Population III binary}

The binary system under study consists of a star of Population III of $50\,M_{\odot}$ and a black hole of $30\,M_{\odot}$. The transfer of mass toward the compact object occurs exclusively by overflowing the Roche lobe. The orbital separation is calculated by approximating the radius of the star by the mean radius of the Roche lobe\cite{Paczynski_1971}. The initial physical properties of the star are adapted from Ref.~\refcite{Marigo_2001}. Considering that the rate of mass accretion occurs on a thermal timescale, that is, the rate of mass loss of the star is regulated by the removal of the thermal equilibrium imposed on the star envelope by the process of mass loss, it is obtained that the black hole accretes matter at a super-Eddington rate.\par

\subsection{Accretion disk}

In an accretion regime super-Eddington, the disks are optically and geometrically thick. This type of disc has been studied by several authors (Refs.~\citen{SS_1973,Paczynski_Wiita_1980,Jaroszynski_1980,Abramowicz_1980,Abramowicz_1988,Beloborodov_1998,Fukue_2004,Akizuki_2006}).  In this work we model the accretion disk considering a critical disk dominated by advection and demagnetized as proposed in Ref.~\refcite{Fukue_2004}, and a disk dominated by advection with toroidal magnetic fields as proposed in Ref.~\refcite{Akizuki_2006}. For distances to the compact object of less than 100 gravitational radius we adopt model of magnetized accretion disk, while for larger distances the first is considered model of demagnetized accretion disk. The free parameters of the model are the angular momentum transport $\alpha$, the beta factor of the plasma $\beta$, and the disk advection parameter $f$.

\subsection{Jets}

We consider that in a region $r < 100\,r_{\mathrm{g}}$ of the accretion disk the ejected matter is collimated forming relativistic jets. Colimation can be explained by the radiatively driven wind ejected at $r > 100\,r_{\mathrm{g}}$ and by the magnetic field ejected from the disk (See Refs.~\citen{Akizuki_2006,Takeuchi_2010}). The launching mechanism of the jet is magnetohydrodynamic. The magnetic field in the launching region requires equipartition between the kinetic and magnetic energy density. The acceleration of particles takes place in a compact region of the jet in which the kinetic energy density dominates over the magnetic energy density. We assume that the relativistic particle injection function is a power law in the energy of the particles:

\begin{equation}
Q(E,z)=Q_{0}\frac{E^{-\mathrm{p}}}{z}\qquad\left[Q\right]=\mathrm{erg^{-1}s^{-1}cm^{-3}}.
\end{equation}

The spectral index taken at the standard  value is p = 2. The normalization constant Q0 for each type of particle depends on the fraction of the kinetic power of the jet used to accelerate particles and the fraction shared between hadrons and leptons in the jets\cite{Romero_MQp_2008}.\par
The radiative processes in the jets are electronic and proton synchrotron, synchrotron self-Compton, relativistic Bremsstrahlung, proton-proton inelastic collisions, and decay of neutral pions by photohadronics interactions. For the calculations see Refs.~\citen{Blumenthal_1970,Mannheim_1994,Atoyan_2003,Kelner_2006,Bosch-Ramon_2006,Kelner_2008,Romero_MQp_2008,Romero_Corona_2010,Vila_2010,Reynoso_2011}. The luminosities produced by each radiative process are calculated in the co-moving reference system with the jet, where the particle distributions are isotropic. The spectral energy distributions (SED) in the observer's reference system are obtained by applying the appropriate Lorentz transformations. The total SED produced in the jets is modified by several absorption processes. In this work we have studied absorption of gamma photons produced in the jets: with the UV photons of the stellar radiation field (external absorption) and with the synchrotron photons produced inside the jets (internal absorption), for the calculations we followed Refs.~\citen{Dubus_2006,Cerutti_2009,Romero_absorption_2010}. The main parameters of the jet are listed in Table 1.

\begin{table}[ph]
\tbl{Model parameters.}
{\begin{tabular}{@{}cccc@{}} \toprule
Parameter & Symbol & Value & Unit \\ \colrule
disk luminosity & $L_{\mathrm{disk}}$ & $10^{40}$ &  $\mathrm{erg\,s^{-1}}$ \\
jet kinetic luminosity in $z_{0}$ & $L_{\mathrm{jet}}$ & $10^{41}$ &  $\mathrm{erg\,s^{-1}}$ \\
jet's content of relativistic particles & $q_{\mathrm{jet}}$ & $0.1$ \\
jet's bulk Lorentz factor $z_{0}$ &  $\Gamma_{\mathrm{jet}}$ & $1.7$\\
jet semi-opening angle tangent & $\chi$ & $0.1$ \\ 
gravitational radius & $r_{g}$ & $44.3$ & $\mathrm{Km}$ \\ 
jet's launching point & $z_{0}$ & $100$ & $r_{g}$ \\
size of acceleration region & $\Delta z$ & $200$ & $r_{g}$ \\
magnetic field in $z_{0}$ & $B(z_{\mathrm{0}})$ & 1.1 $\times 10^{7}$ & $\mathrm{G}$ \\ 
particle injection spectral index & $p$ & $2.0$\\
acceleration efficiency & $\eta$ & 0.1\\
hadron-to-lepton energy ratio & $a$ & 0.1\\ \botrule
\end{tabular} \label{ta1}}
\end{table}

\section{Results}

Figs. 1 and 2 show the distribution of thickness and temperature in the accretion disk as a function of the distance to the compact object, for different values ​​of the angular momentum transport efficiency. In all cases, the beta factor of the plasma was set to $0.5$, corresponding to a strict equiparticon. The disk advection parameter is $0.5$, which is a reasonable value for super-Eddington accretion discs\cite{Fukue_2004,Akizuki_2006}. Fig. 3 shows the SED of the accretion disk. In this case, we have varied the transport momentum of angular momentum within physically permitted values\cite{Fukue_2004}. 

\begin{figure}[!t]
  \centering
  \includegraphics[width=0.75\textwidth]{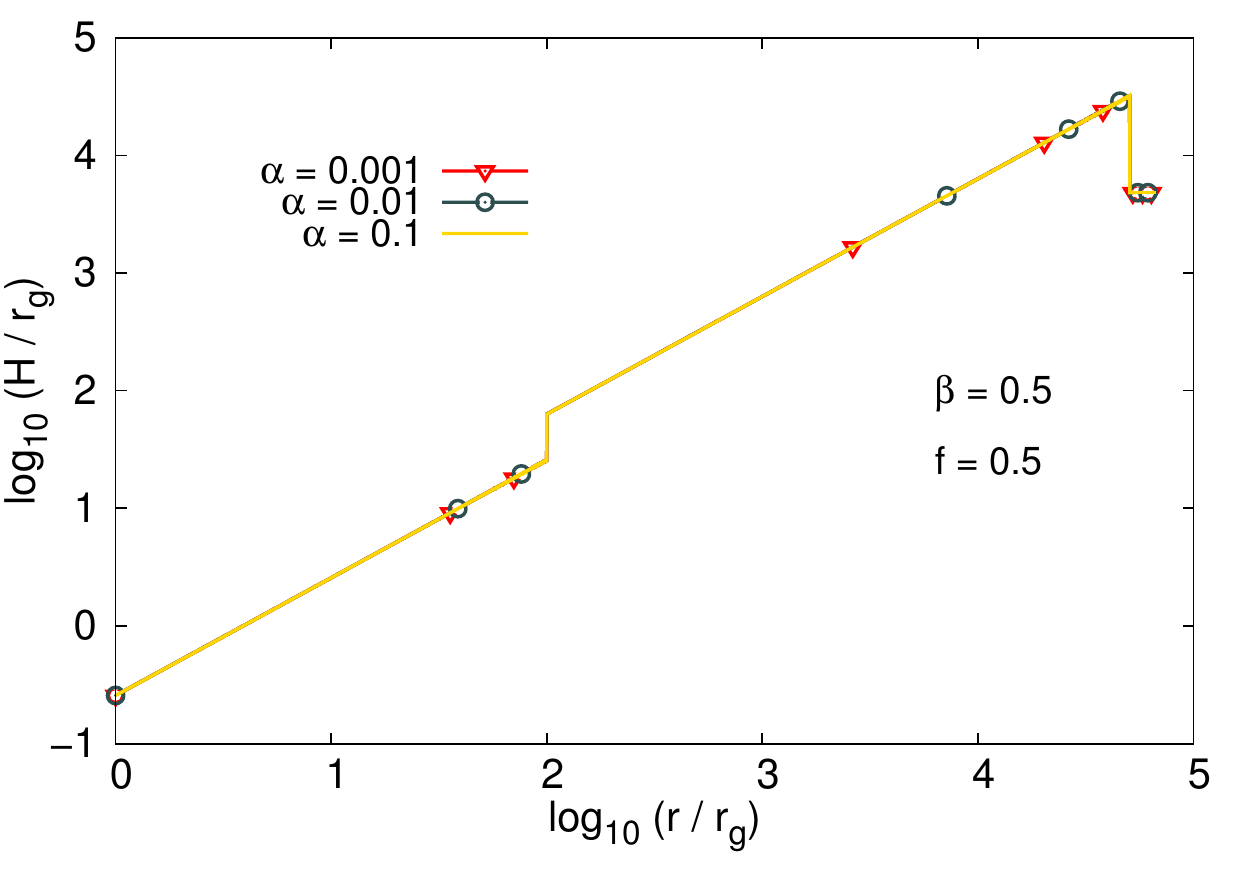}
  \caption{Thickness of the accretion disk.
}
  \label{F_espesor}
\end{figure}
\begin{figure}[!t]
  \centering
  \includegraphics[width=0.75\textwidth]{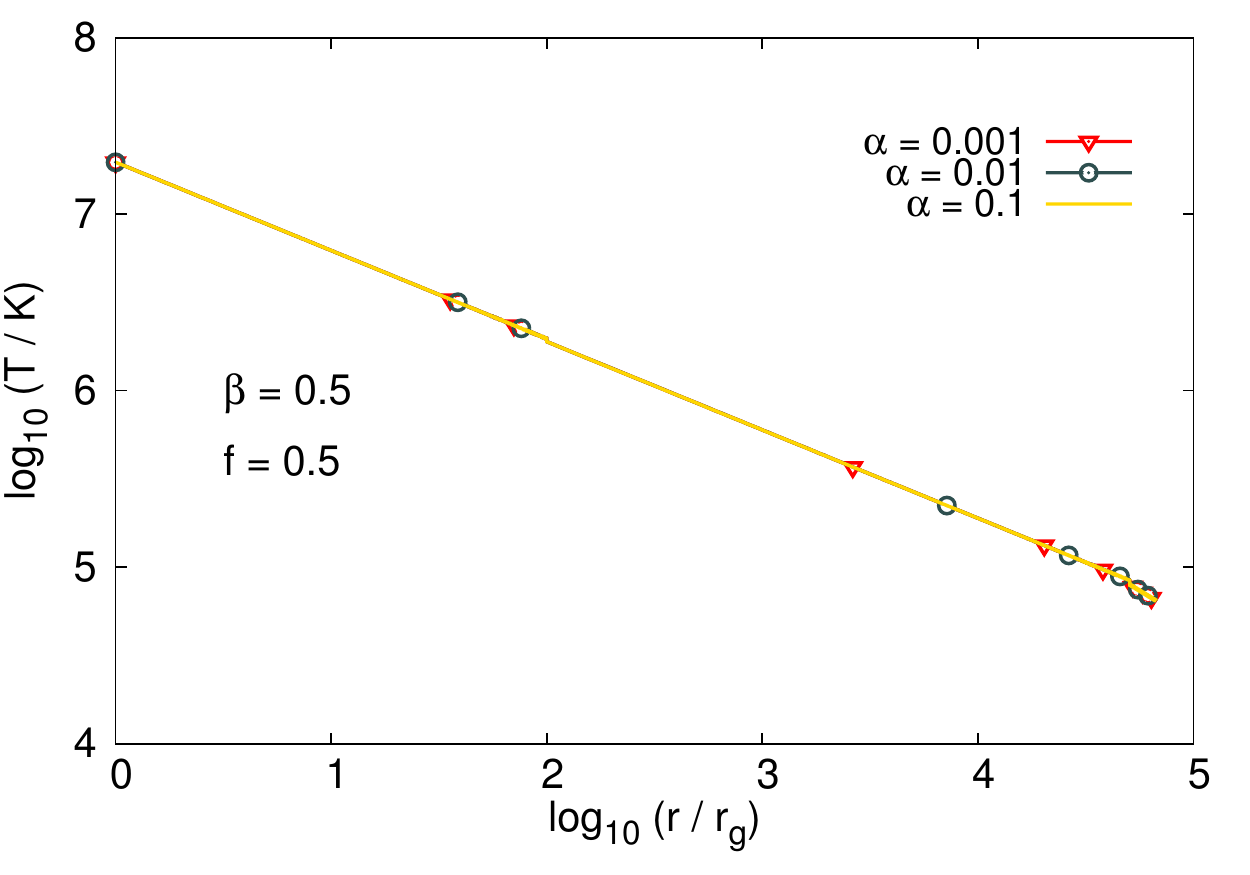}
  \caption{Temperature of the accretion disk
}
  \label{F_temperatura}
\end{figure}
\begin{figure}[!t]
  \centering
  \includegraphics[width=0.75\textwidth]{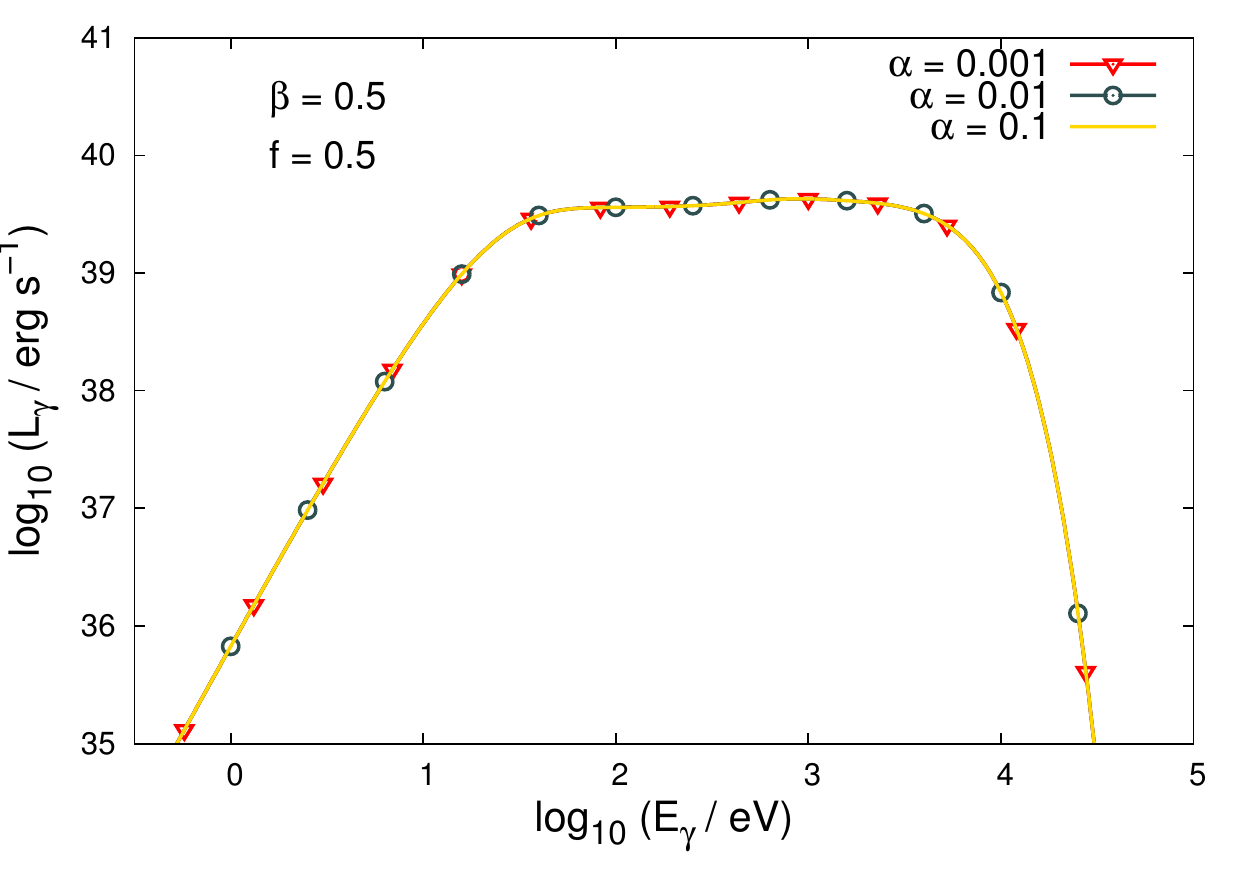}
  \caption{Spectral energy distribution of the accretion disk.
}
  \label{F_SEDdisco}
\end{figure}

Fig. 4 shows the spectra produced in the radiative processes considered here. The main contributions to the SED are synchrotron radiation, inverse Compton and $p\gamma$ interactions. The relative importance between the hadronic and leptonic species is $a=L_{\mathrm{p}}/L_{\mathrm{e}}=0.1$. We have considered that the particle acceleration process is very efficient ($\eta=0.1$), so there is a high energy cutoff in the SEDs. In Fig. 5, we show the opacity map produced by the photon absorption in the stellar field, in the relevant energy range and along the complete orbit ($\phi$ is the orbital phase in units of $2\pi$; we note that $\phi=0=1$ corresponds to the compact object at opposition, i.e. superior conjunction). So, of the two absorption processes taken into account, the internal absorption is the most relevant, suppressing completely the emission produced in the jets for energies greater than a few MeV. This can be seen in Fig. 6. 

\begin{figure}[!t]
  \centering
  \includegraphics[width=0.75\textwidth]{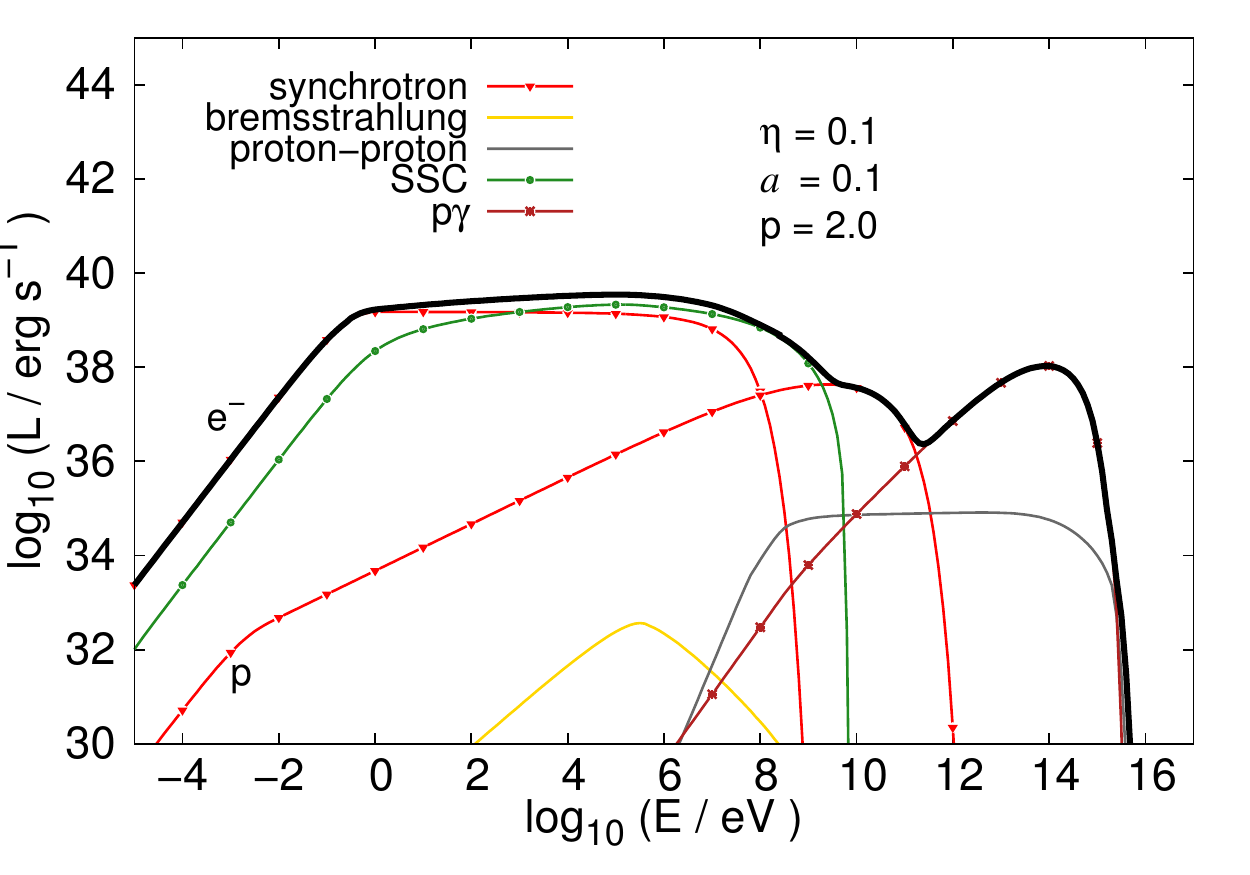}
  \caption{Spectral energy distribution of the relativistic particles in the jets.
}
  \label{F_SEDjet}
\end{figure}
\begin{figure}[!t]
  \centering
  \includegraphics[width=0.95\textwidth]{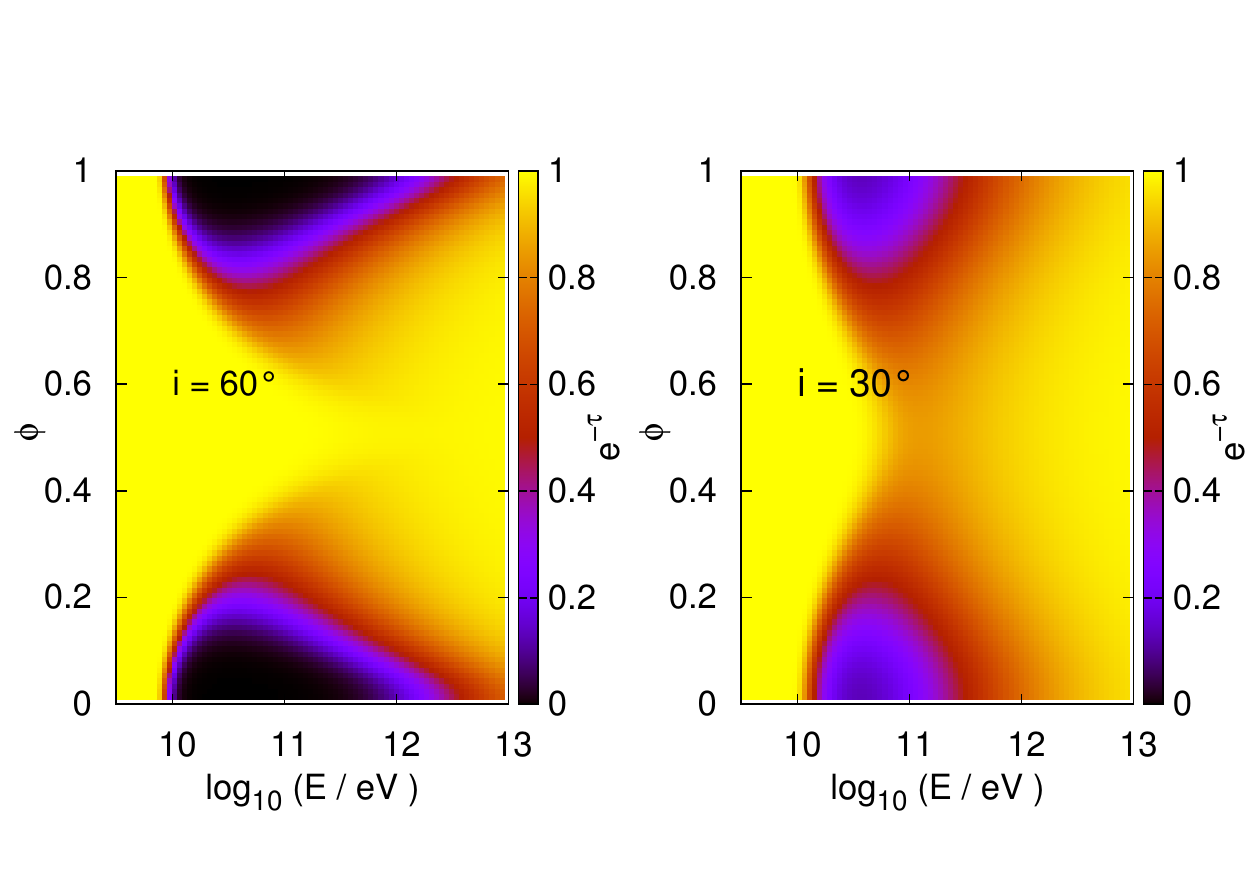}
  \caption{Map of the absorption produced by the stellar photon field for two viewing angles.
}
  \label{F_map}
\end{figure}
\begin{figure}[!t]
  \centering
  \includegraphics[width=0.75\textwidth]{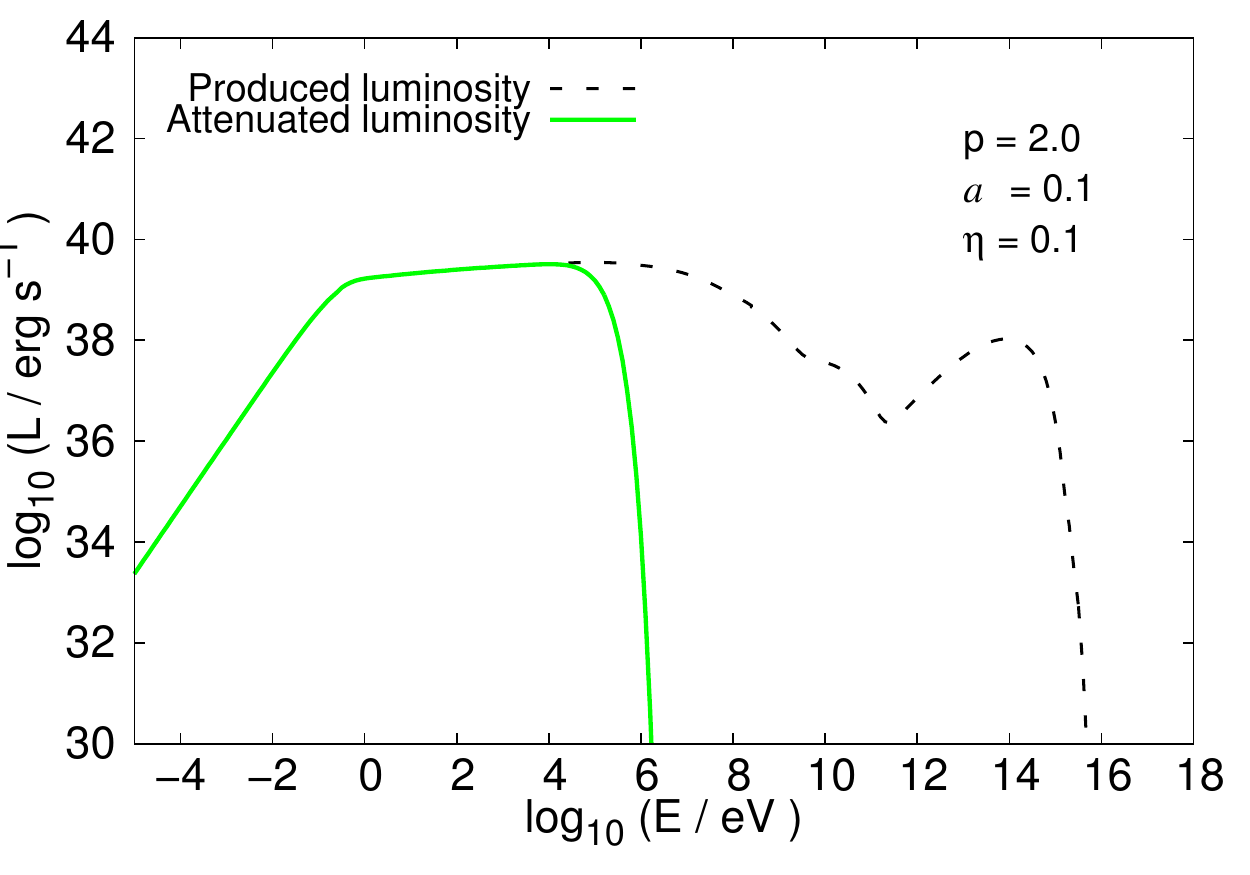}
  \caption{Spectral energy distribution in the jets attenuated by internal absorption.
}
  \label{F_SEDcorr}
\end{figure}

\section{Discussion and summary}

We have developed a simple model of microquasar of Population III. We determined that accretion disks differ significantly with respect to
those of the galactic microquasars that have been detected so far (except SS433, see Ref.~\refcite{Reynoso_SS433_2008}). These disks would have produced an important radiation in the X-ray band, which would be attenuated by direct Compton interactions with the disc's wind particles. In the jets an intense emission of gamma rays is produced, which is suppressed internally by photon-photon interactions, resulting in the production of relatively low energy pairs. These pairs were cooled mainly by synchrotron radiation, so electromagnetic cascades do not occur.
In the future, we will include in the model a more detailed study of the radiative impact of the wind of the accretion disk. The contribution of the pairs produced in the $pp$ and $p\gamma$ interactions would also be included in a forthcoming work.




\begin{thebibliography}{0}    

\bibitem{Mirabel_BH_2011} Mirabel, I. F., Dijkstra, M., Laurent, P., Loeb, A., \& Pritchard, J.~R.\ 2011, Astronomy \& Astrophysics, 528, A149.

\bibitem{Tueros_CR_2014} Tueros, M., del Valle, M.V., \& Romero, G.E.\ 2014,  Astronomy \& Astrophysics, 570, L3.

\bibitem{Paczynski_1971} Paczy{\'n}ski, B.\ 1971, Annual Review of Astronomy and Astrophysics, 9, 183.

\bibitem{Marigo_2001} Marigo, P., Girardi, L., Chiosi, C., \& Wood, P.~R.\ 2001, Astronomy \& Astrophysics, 371, 152.

\bibitem{SS_1973} Shakura, N.~I., \& Sunyaev, R.~A.\ 1973, Astronomy \& Astrophysics, 24, 337.

\bibitem{Paczynski_Wiita_1980} Paczy{\'n}sky, B., \& Wiita, P.~J.\ 1980,  Astronomy \& Astrophysics, 88, 23.

\bibitem{Jaroszynski_1980} Jaroszynski, M., Abramowicz, M.~A., \& Paczynski, B.\ 1980, Acta Astronomica, 30, 1.

\bibitem{Abramowicz_1980} Abramowicz, M.~A., Calvani, M., \& Nobili, L.\ 1980, The Astrophysical Journal, 242, 772.

\bibitem{Abramowicz_1988} Abramowicz, M.~A., Czerny, B., Lasota, J.~P., \& Szuszkiewicz, E.\ 1988, The Astrophysical Journal, 332, 646.

\bibitem{Beloborodov_1998} Beloborodov, A.~M.\ 1998, Monthly Notices of the Royal Astronomical Society, 297, 739.

\bibitem{Fukue_2004} Fukue, J.\ 2004, Publications of the Astronomical Society of Japan, 56, 569.

\bibitem{Akizuki_2006} Akizuki, C., \& Fukue, J.\ 2006, Publications of the Astronomical Society of Japan, 58, 469.

\bibitem{Takeuchi_2010} Takeuchi, S., Ohsuga, K., \& Mineshige, S.\ 2010, Publications of the Astronomical Society of Japan, 62, L43. 

\bibitem{Romero_MQp_2008} Romero, G.~E., \& Vila, G.~S.\ 2008, Astronomy \& Astrophysics, 485, 623.

\bibitem{Blumenthal_1970} Blumenthal, G.~R., \& Gould, R.~J.\ 1970, Reviews of Modern Physics, 42, 237. 

\bibitem{Mannheim_1994} Mannheim, K., \& Schlickeiser, R.\ 1994, Astronomy \& Astrophysics, 286, 983.

\bibitem{Atoyan_2003} Atoyan, A.~M., \& Dermer, C.~D.\ 2003, The Astrophysical Journal, 586, 79.

\bibitem{Kelner_2006} Kelner, S.~R., Aharonian, F.~A., \& Bugayov, V.~V.\ 2006, Physical Review D, 74, 034018.

\bibitem{Bosch-Ramon_2006} Bosch-Ramon, V., Romero, G.~E., \& Paredes, J.~M.\ 2006, Astronomy \& Astrophysics, 447, 263. 

\bibitem{Kelner_2008} Kelner, S.~R., \& Aharonian, F.~A.\ 2008, Physical Review D, 78, 034013.

\bibitem{Reynoso_2011} Reynoso, M.~M., Medina, M.~C., \& Romero, G.~E.\ 2011, Astronomy \& Astrophysics, 531, A30.

\bibitem{Romero_Corona_2010} Romero, G.~E., Vieyro, F.~L., \& Vila, G.~S.\ 2010, Astronomy \& Astrophysics, 519, A109. 

\bibitem{Vila_2010} Vila, G.~S., \& Romero, G.~E.\ 2010, Monthly Notices of the Royal Astronomical Society, 403, 1457.

\bibitem{Dubus_2006} Dubus, G.\ 2006, Astronomy \& Astrophysics, 451, 9. 

\bibitem{Cerutti_2009} Cerutti, B., Dubus, G., \& Henri, G.\ 2009, Astronomy \& Astrophysics, 507, 1217. 

\bibitem{Romero_absorption_2010} Romero, G.~E., Del Valle, M.~V., \& Orellana, M.\ 2010, Astronomy \& Astrophysics, 518, A12. 

\bibitem{Reynoso_SS433_2008} Reynoso, M.~M., Romero, G.~E., \& Christiansen, H.~R.\ 2008, Monthly Notices of the Royal Astronomical Society, 387, 1745. 


\end{thebibliography}
\end{document}